\documentclass[prd,preprint,aps,showpacs]{revtex4}
\usepackage{latexsym}
\usepackage{amsmath}
\usepackage{amssymb}
\usepackage[all]{xy}

\newcommand{\be}{\begin{equation}}
\newcommand{\ee}{\end{equation}}
\newcommand{\bn}{\begin{eqnarray}}
\newcommand{\en}{\end{eqnarray}}
\newcommand{\p}{\partial}
\newcommand{\pslash}{p \hspace{-1.7mm} /}

\begin{document}

\title{Induced deformation of the canonical structure and UV/IR duality in $(1+1)D$}
\author{L. S. Grigorio$^{a}$, M. S. Guimaraes$^{a}$, C. Wotzasek$^{a}$}
\affiliation{$\mbox{}^{a}$Instituto de F\'\i sica, Universidade Federal do Rio de
Janeiro, 21945, Rio de Janeiro, Brazil\\
{\sf E-mail: clovis@if.ufrj.br}\\
\today}


\begin{abstract}

The purpose of this work is two fold. Working in the framework of $(1+1)D$ Lorentz
violating field theories we will investigate in the first place the general claim
that fermionic interactions may be equivalent to a deformation of the canonical
structure of the theory. Second the deformed theory will be studied using duality
reasoning to address the behavior of the Infra-Red and Ultra-Violet regimes.

\end{abstract}

\maketitle


\section{Introduction and motivations}

It is known as a general fact that the presence of fermions in a theory can in many
instances be effectively accounted for by a deformation in the canonical structure
of the theory without fermions \cite{carmona}. This can be traced to the possible
induction of non trivial terms due to quantum fermionic fluctuations \cite{redlich}
which changes the symplectic structure. In this work we want to address this
important fact in the context of Lorentz violating theories \cite{testsLV,
mattingly}. In this context our work has a two fold purpose: in the first place, we
want to strength the knowledge of the fact that Lorentz violation as described by
the deformation of the canonical structure, a sometimes \emph{ad hoc} imposition,
may in fact have its origin in Lorentz violating fermionic interactions. This will
be done by working out an explicit example of a scalar field interacting with
fermions in $(1+1)D$. The effective scalar theory with the quantum fermionic
fluctuations accounted for is obtained considering the Goldstone-Wilczek mechanism.
This is a new and important result as it provides another non trivial example of
generation of deformed structures induced by interactions. As a second objective we
will study the resulting deformed theory under the scope of duality. We want to shed
some light on the physical mechanism involved in the canonical structure deformation
of the dual theory and, in particular, we want to gain some insights on the
relations between the IR and UV scales in a Lorentz violating background. This is an
important inquire since it is generally demanded that Lorentz violation effects
takes place in nearly unobservable scales so that its effects could have managed to
remain undetected so far, but if duality could provide us with a map that connects
this scale to observable ones the safety of this argument would be spoiled.

It is important to briefly review some results on this subject in order to set the
ground on which this work relies. The recent surge in the study of Lorentz violation
has its roots in the exploration of physics beyond the standard models of particle
physics and cosmology. As a matter of fact it is generally believed that Lorentz
invariance cannot hold at such extreme regions as the Planck scale and it is
expected that some relics of this violation are translated on to observable effects
at accessible scales. Thus a reasonable way to approach the problem is to understand
what kind of phenomena signaling a departure of the Lorentz invariance we should
expect to see. This is the rationale behind the proposed extended standard model
(ESM) \cite{colkostESM}, for example, which catalogues the possible Lorentz
violating terms that can be added to the conventional standard model lagrangian of
particle physics.

Along the same lines, and in fact closely related, is the study of non-commutative
theories. Initially proposed by Heisenberg and by Snyder \cite{snyder} it gained
much impulse recently due to its natural appearance in distinct arenas such as the
quantum Hall phenomena and in the context of string theory, signaling that such
structures might have indeed a fundamental origin. There are two main branches
characterizing non-commutative theories: one is the study of spacetime
non-commutativity defined by having nonvanishing commutators among the coordinates
of spacetime leading to novel field theory models. The other is the noncommutativity
of field operators imposed by a deformation in the canonical structure of the field
theories \cite{carmona}. Obviously both approaches may lead to theories in which
Lorentz invariance can be broken. In fact there seems to be a remarkable unity in
all approaches to Lorentz violation, the ESM Lorentz violating terms may be
connected to noncommutativity of spacetime \cite{carrollkostelecky} and on the other
hand can also be induced by deformation of the canonical structure
\cite{gamboasarrion}. Those results builds up the case that consistent theories may
be constructed giving up Lorentz invariance without diving in to such unacceptable
consequences as causal and unitarity violations.

In this work our interest relies primarily in the study of deformed canonical
structures as induced by quantum dynamical effects. This is a very important
phenomenon and its occurrence reveals non trivial structures. For example in
$(2+1)D$ it is known that if the Maxwell theory is written in interaction with
massive fermions the Chern-Simons term cannot be disregarded since it will be
induced by the fermionic dynamics anyway\cite{redlich}. The resulting
Maxwell-Chern-Simons theory shows non trivial topological properties that are not
evident initially. This is a particular case of a fundamental result. There are
other examples such as the induced Berry phases represented by Wess-Zumino terms
which has been shown recently to have a decisive role in determining the quantum
phase transitions of magnetic systems \cite{sachdev}. More akin to our work is the
result obtained in \cite{gamboasarrion} which is the generalization to a $(3+1)D$
Lorentz violating framework of the $(2+1)D$ case discussed above. It was shown that
the Carroll-Field-Jackiw model \cite{cfj} can be written as the conventional Maxwell
theory with deformed canonical relations. This result is to be related to that
discussed by Kostelecky and Jackiw \cite{jk} which investigates the induction of the
Lorentz violating Chern-Simons-like term defining the Carroll-Field-Jackiw model by
the quantum fluctuations of a Lorentz violating fermionic dynamics. In the present
work we will investigate a similar result in $(1+1)D$ concerning scalar fields
interacting with fermions and where the fermions will be dealt with by the
Goldstone-Wilczek mechanism \cite{gw}.

The work is organized as follows: in the next section our results will be presented.
We will start by defining the effective scalar theory originating from taking into
account the Lorentz violating fermionic interaction as dictated by the
Goldstone-Wilczek mechanism \cite{gw}. We will proceed with an analysis of the
propagating modes highlighting the effects of the Lorentz violation. After a
discussion of the symplectic structure of the effective theory, that sets straight
our expectations concerning the canonical relations in a dual picture, we will seek
this dual formulation to find that a possible connection can be established between
the infra-red and ultra-violet regimes. We will close with the concluding remarks
where it is discussed the potencial consequences of this IR/UV mapping and our
concerns regarding its limitations. An appendix is included to further discuss the
derivation of the induced current in the Goldstone-Wilczek mechanism.

\section{Results}

Before delving into the actual presentation of our results it will pay off to
elaborate a bit further on the framework of noncommutative fields. As mentioned
above, there are two major trends involving noncommutativity. Without a doubt the
great majority of works in this field is dedicated to space-time noncommutativity
embodied in relations such as
\begin{eqnarray}
\label{c01}
 [x_\mu, x_\nu] = i \theta_{\mu\nu}
\end{eqnarray}
where $x_\mu$ are coordinates of space-time and $\theta_{\mu \nu}$ is an
antisymmetric constant tensor with dimension $(length)^2$. This kind of structure
shows up in a variety of contexts, most notably in string theory where the
coordinates represent longitudinal directions of D-branes, as seen by the ends of
open strings, in the presence of a B-field background, and also in the quantum Hall
effect where the presence of a strong magnetic field induces a noncommutativity
among the coordinates of the particle in a plane. However here the noncommutativity
among the coordinates comes from the projection of the operators over the Lowest
Landau Level.

But it is the other trend which is the one directly related to our results \cite{carmona}. The
quantum theory of noncommutative fields has been elaborated as a generalization of
noncommutative quantum mechanics which are rather different from the usual quantum
field theory over a canonical noncommutative space-time \cite{douglas}. The basic idea of this
procedure is to include minimal modifications to the canonical structure of the
field theory, or, equivalently, of its symplectic structure \cite{carmona},
amounting to adding a tiny violation of the microcausality principle. In these
works, the notion of noncommutative fields are introduced by
\begin{eqnarray}
\label{c02}
\left[\Phi_i(x), \Phi_j(y)\right] = i \varepsilon_{ijk} B^k \delta^{(3)}(x-y)\nonumber\\
\left[\Phi_i(x), \Pi_j(y)\right] = i \delta_{ij} \delta^{(3)}(x-y) \nonumber\\
\left[\Pi_i(x), \Pi_j(y)\right] = i \varepsilon_{ijk} \Theta^k \delta^{(3)}(x-y)
\end{eqnarray}
where $ i, j = 1, 2 , 3$ and $\Pi_i$ are the conjugate momenta. We mention that, due
to the presence of $\delta^{(3)}(x - y)$ in the right-hand side, the constant
vectors $B_k$ and $\Theta_k$ have canonical dimension of length and mass,
respectively introducing an ultra-violet and an infra-red scale respectively.

It is important to point out that Lorentz symmetry violation due to
noncommutativity of fields is not yet established as compared to the usual
formulation of field theory over canonical noncommutative spacetime simply because
we still lack explicit model realizations. In a recent paper the noncommutative
field space formulation was used to analyze the abelian bosonization for a two
dimensional system \cite{Das:2004vc}. The rationale there was that an analysis in a D = 2
spacetime theory can be useful in disclosing the basic physics underlying this
problem. They found that for chiral bosons in a noncommutative field space conformal
invariance continues to hold and that the non-commutativity in the field space leads
to free fermions when chiral bosons are fermionized.

In a recent report, the connection between Lorentz invariance violation and
noncommutativity of fields in a quantum field theory of chiral bosons was resumed
with the investigation of a generalized model of non-commutative field space chiral
bosons with a real one-parameter deformed symplectic algebra \cite{Abreu:2004xe} which was
investigated upon the soldering of the individual chiralities.

In the present work we want to study the possibility of having the low-energy sector
of an effective real scalar field model having its canonical structure deformed by
quantum fluctuations of a fermionic field coupled to this scalar field, such that
\be [\Pi(x), \Pi(y)]\to l(x-y) \ee with $l(x-y)$ an anti-symmetry form to be
determined below, while the other brackets remain unchanged.


\subsection{The Goldstone-Wilczek mechanism}

Here we shall follow the strategy of \cite{colkostESM} and consider
bosonic/fermionic model with Lorentz violating interaction. Consider the following
Lorentz violating action:
\begin{eqnarray}
\label{m01}
  S = \int d^2x \left( \frac 12 \partial_{\mu}\phi\partial^{\mu}\phi +
  i\overline{\psi}\gamma^{\mu}\partial_{\mu}\psi - \theta \phi'\overline{\psi}\gamma^{1}\psi - g\overline{\psi}e^{\gamma_5\phi}\psi\right)
\end{eqnarray}
it describes a scalar field $\phi$ coupled with a massless fermion field $\psi$.
Here $\gamma^0 = \sigma^1$, $\gamma^1 = i\sigma^3$ and $\gamma_5 = i\sigma^2$. The
derivative interaction explicitly violates Lorentz invariance as it defines a
constant tensor which selects a preferred Lorentz frame:
\begin{eqnarray}
\label{m02}
 \phi'\overline{\psi}\gamma^{1}\psi =
P^{\mu\nu}(\partial_{\mu}\phi)\overline{\psi}\gamma_{\nu}\psi ; \;\;\;\; P^{\mu\nu}
=\left(\begin{array}{cc}
0&0\\
0 & \theta
\end{array}\right).
\end{eqnarray}
We want to consider a framework in which the space-time variations of the scalar
field may be neglected in a first approximation. Ultimately we are searching for an
effective theory describing the low momenta excitations of the scalar field. We also
demand the Lorentz violating effects to be small thus retaining only first order
terms in $\theta$. To construct such an effective theory we must take into account
the contribution of the fermionic fluctuations defining the effective action by the
expression:
\begin{eqnarray}
\label{m03}
 e^{iS_{eff}}= \int {\cal{D}\overline{\psi}\cal{D}\psi} e^{iS}.
\end{eqnarray}
Further, neglecting higher order scalar derivatives we can write
\begin{eqnarray}
\label{m04}
 e^{iS_{eff}} = e^{\int d^2x  \frac 12 \partial_{\mu}\phi\partial^{\mu}\phi} <e^{-\int
 d^2x\theta \phi'\overline{\psi}\gamma^{1}\psi}> = e^{\int d^2x  \frac 12 \partial_{\mu}\phi\partial^{\mu}\phi} e^{-\int
 d^2x\theta \phi'<\overline{\psi}\gamma^{1}\psi>}
\end{eqnarray}
where
\begin{eqnarray}
\label{m05}
 <\overline{\psi}\gamma^{\mu}\psi> = \int {\cal{D}\overline{\psi}\cal{D}\psi}
 (\overline{\psi}\gamma^{\mu}\psi)
 e^{i\int d^2x \left( i\overline{\psi}\gamma^{\mu}\partial_{\mu}\psi - g\overline{\psi}e^{\gamma_5\phi}\psi\right)}
\end{eqnarray}
is the induced fermion current due to the non-derivative interaction given by the
last term in (\ref{m01}). This expression neglects higher $\theta$ order
contributions and higher $\phi$ derivatives also. The physical meaning is that we
are supposing a classical behavior for the fermionic current at the scale of
variations of $\phi$, that is: $<(j^{\mu} -<j^{\mu}>)^2> = 0$. It is straightforward
to calculate this induced current \cite{gw, af} imposing the condition $|\partial
\phi| \ll |g|$, it is given by
\begin{eqnarray}
\label{m06}
 <\overline{\psi}\gamma^{\mu}\psi> = - \frac{1}{2\pi} \varepsilon^{\mu\nu}\partial_{\nu}
 \phi.
\end{eqnarray}
For a somewhat more detailed discussion of these matters and a bosonized view see
the appendix. Then the effective Lorentz violating action is given by
\begin{eqnarray}
\label{m07} S_{eff} =  \int d^2x \left( \frac 12
\partial_{\mu}\phi\partial^{\mu}\phi - \frac{\theta}{2\pi} \phi'\dot{\phi} \right)
\end{eqnarray}
where $\phi' \equiv \partial_1 \phi$ and $\dot{\phi} \equiv \partial_0 \phi$.
Observe that it does not depend on the coupling $g$, nevertheless the scale of
validity of our approximation, and therefore of the effective theory itself, is
defined by $g$, which has mass dimension $1$ setting an ultraviolet cutoff for the
scalar momentum in the effective theory. This observation will be very important in
the discussion of the dual formulation and the UV/IR dual map.


\subsection{Modes decomposition}

The violation of the Lorentz invariance has a very interesting manifestation here.
The effective action (\ref{m07}) describes two independent propagating modes with
different velocities. This can be explicitly seen by making a chiral decomposition
as follows. Let us write (\ref{m07}) as:
\begin{eqnarray}
\label{m08} S_{eff} =  \int d^2x \left(
\Pi\dot{\phi}- \frac 12 \Pi^2 - \frac 12 \phi'^2- \frac{\theta}{2\pi} \phi'\dot{\phi} \right).
\end{eqnarray}
where $\Pi$ is an auxiliary field. We can further redefine $\Pi \rightarrow \eta + \frac{\theta}{2\pi}\phi'$ giving
\begin{eqnarray}
\label{m09} S_{eff} =  \int d^2x \left(
\eta\dot{\phi}- \frac 12 \left( \eta + \frac{\theta}{2\pi}\phi' \right)^2 - \frac 12 \phi'^2\right).
\end{eqnarray}
Now the following field redefinition can be made
\begin{eqnarray}
\label{m10} \phi &=& \phi_{+}  +  \phi_{-}\nonumber\\
  \eta &=& u \left( \phi'_{+}  -  \phi'_{-}\right)
\end{eqnarray}
where $u = \sqrt{1+\left(\frac{\theta}{2\pi}\right)^2}$. With this we finally obtain the result
\begin{eqnarray}
\label{m11} S_{eff} =  S_{+} + S_{-},
\end{eqnarray}
where
\begin{eqnarray}
\label{m12} S_{+} &=& \int d^2x \left(u\dot{\phi}_{+}\phi'_{+} - \frac 12 \left[ 1 +  \left(u + \frac{\theta}{2\pi} \right)^2 \right] \phi'^2_{+}\right)  \nonumber\\
  S_{-} &=& \int d^2x \left(-u\dot{\phi}_{-}\phi'_{-} - \frac 12 \left[ 1 +  \left(u - \frac{\theta}{2\pi} \right)^2 \right] \phi'^2_{-}\right).
\end{eqnarray}
We immediately see that the velocities of each mode are different. In fact, for the sensible case $\theta\ll 1$, we have:
\begin{eqnarray}
\label{m13} v_{+} &=&  \left(1 + \frac{\theta}{2\pi} \right) \nonumber\\
  v_{-} &=& \left(1 - \frac{\theta}{2\pi} \right).
\end{eqnarray}
Of course, this could be inferred directly from the dispersion relation following
from the original effective action (\ref{m07}), but it is instructive to reveal the
possibility of factorization of the modes as depicted in (\ref{m12}). This is to be
compared to the well known factorization of the Proca model in $(2+1)D$ in its
self-dual components \cite{Banerjee:1998xv}.

\subsection{Symplectic structure}

From the symplectic matrix one can read the Poisson brackets of the model. These are
not the canonical ones. But the action (\ref{m07}) and the free scalar theory with
non-canonical brackets both lead to the same equations of motion.

More quantitatively, we can use the reduced order form (\ref{m08}) again from which
follows immediately the inverse symplectic matrix \be \label{d2} f= \left(
\begin{array}{cc}
    0 & -1 \\
    1 & -\frac{\theta}{\pi} \p_x\\
    \end{array}\right)\delta(x-y)
\ee which shows that the bracket $\{\Pi(x),\Pi(y)\}$ is deformed. It is then easy to
verify that the free scalar Hamiltonian \be \label{d3} H=\int dx \frac 1 2 \left[
\Pi^2 + (\phi ')^2\right] \ee with the brackets given by \bn \label{d4}
&& \{\phi(x),\phi(y)\}=0 \nonumber\\
&& \{\phi(x),\Pi(y)\}=\delta(x-y) \nonumber \\
&& \{\Pi(x),\Pi(y)\} = \frac{\theta}{\pi} \p_x \delta(x-y) \en lead to the same
equations of motion obtained by minimizing the action (\ref{m07}). Thus we may
interpret the Lorentz violating induced term due to fermionic interaction as a
modification of the canonical Poisson brackets of the free field theory.

Observe that the deformation showed up in the momentum sector. The parameter
$\theta$ is dimensionless but the deformation is proportional to the derivative and
thus has mass dimension $1$, fitting the general discussion bellow (\ref{c02}).
Incidentally this is interpreted as an infra-red deformation (see the comments on
\cite{falomir0504032}). Since duality has as a general property the interchange of
potential and kinetic contributions, could it be possible to transfer this
deformation to the ultra-violet sector by duality transformations? In fact we will
show in the next section that an exact dual representation exists with brackets
given by \bn \label{d6}
&& \{\Sigma(x),\Sigma(y)\}=\frac{\theta}{2\pi} \epsilon(x-y) \nonumber\\
&& \{\Sigma(x),P(y)\}=\delta(x-y) \nonumber \\
&& \{P(x),P(y)\} = 0 \en where the escalar field $\Sigma$ is the dual representation of the
$\phi$ field and $\epsilon(x-y)$ is the skew symmetric step function with the
property $\partial_x\epsilon(x-y) = 2\delta(x-y)$. This would stand for a
ultra-violet deformation. We will discuss these remarks more precisely in the
following section.

\subsection{The dual formulation}

We will now seek the dual formulation of the theory discussed so far. Observe that
the action (\ref{m07}) can be cast in the form:
\begin{eqnarray}
\label{m14} S_{eff} =  \int d^2x  \frac 12
\partial_{\mu}\phi M^{\mu\nu}\partial_{\nu}\phi
\end{eqnarray}
where
\begin{eqnarray}
\label{m15}
 M^{\mu\nu}
&=&\left(\begin{array}{cc}
1&\frac{\theta}{2\pi}\\
\frac{\theta}{2\pi} & -1
\end{array}\right) = \left(1 + \left(\frac{\theta}{2\pi}\right)^2\right)M^{-1}_{\mu\nu},
\end{eqnarray}
the physics described by (\ref{m14}) is not altered if we introduce an auxiliary
field $\Pi^{\mu}$.
\begin{eqnarray}
\label{m16} S_{eff} \rightarrow  \int d^2x  \left(\Pi^{\mu}\partial_{\mu}\phi -
\frac 12 \Pi_{\mu} (M^{-1})^{\mu\nu}\Pi_{\nu}\right)
\end{eqnarray}
$\Pi^{\mu}$ can be integrated out (in a path integral sense) leading us back to
(\ref{m14}). On the other hand $\phi$ can be viewed as a Lagrange multiplier forcing
the constraint
\begin{eqnarray}
\label{m17} \partial_{\mu}\Pi^{\mu} = 0 \Rightarrow \Pi^{\mu} =
\varepsilon^{\mu\nu}\partial_{\nu}\Sigma,
\end{eqnarray}
where $\varepsilon_{01}=1 \Rightarrow \varepsilon^{01}=-1$. $\Sigma$ is the dual
field and the action in terms of it is the dual action, which in this particular
case is just the original one, that is, the theory is self-dual after a trivial scaling.
\begin{eqnarray}
\label{m18} S_{eff} \rightarrow \ast S_{eff} = \int d^2x \frac{1}{2\left(1 +
\left(\frac{\theta}{2\pi}\right)^2\right)}
\partial_{\mu}\Sigma M^{\mu\nu}\partial_{\nu}\Sigma
\end{eqnarray}
From (\ref{m16}) a map between $\phi$ and $\Sigma$ follows
\begin{eqnarray}
\label{m19} \left(\begin{array}{c}
\dot{\Sigma}\\
\Sigma'\end{array}\right) = R(\theta) \left(\begin{array}{c}
\dot{\phi}\\
\phi'\end{array}\right) = \left(\begin{array}{cc}
\frac{\theta}{2\pi}& -1\\
-1 & \frac{\theta}{2\pi}
\end{array}\right) \left(\begin{array}{c}
\dot{\phi}\\
\phi'\end{array}\right).
\end{eqnarray}
$R(\theta)$ is the dual map, but it is not the most general one that preserves the
equations of motion. Linearity allows us to consider a linear combination including
the trivial map (the identity map)
\begin{eqnarray}
\label{m20} \tilde{R}(\theta) = a\textbf{1} + b R(\theta)
\end{eqnarray}
where $a$ and $b$ are real constants.

This is a good place to comment on what seems to be a general feature of the kind of
duality discussed here. Under a direct application of the map (\ref{m20}) the action
(\ref{m14}) transforms as
\begin{eqnarray}
\label{m21} S_{eff} \rightarrow  \left(\frac{1}{a^2 - b^2\left(1 +
\left(\frac{\theta}{2\pi}\right)^2\right)}\right) S_{eff}
\end{eqnarray}
For $a=0$, $b=1$ the map reduces to the duality map (\ref{m19}), but there is a sign
difference with respect to the dual obtained in (\ref{m18}) through the duality
procedure. This will not affect the equations of motion of course, but it is
nevertheless disturbing and begs for an explanation. The reason for this difference
has its roots in a misuse of the map. For the map is constructed using the equations
of motion of both $\Pi^{\mu}$ and $\phi$ that follows from (\ref{m16}), so it is not
rigourously licit to substitute this on-shell information on the action.
Nevertheless since we are dealing with quadratic actions this sign change is the
only effect of the direct use of the map on the action level, it is only a
reflection of a general property of duality in these cases: it interchanges kinetic
and potential contributions. But care should be taken in more general cases. This in
fact should be all very familiar from Maxwell theory in $(3+1)D$: the analogous
duality amounts to an interchange of electric and magnetic fields ($\vec{E}
\rightarrow \vec{B}$, $\vec{B} \rightarrow -\vec{E}$) which of course changes the
sign of the action when naively applied to it even though the Maxwell theory is
self-dual.

With these warnings in mind we proceed now to the study of the hamiltonian structure
of this theory under the duality map. We are seeking for a dual description in which
the deformed momentum brackets (\ref{d4}) are mapped to the field space deformed
ones (\ref{d6}). The relevant relation is then
\begin{eqnarray}
\label{m22} \{\Sigma'(x),\Sigma'(y)\} &=& \{b\dot{\phi} + \left(a +
b\left(\frac{\theta}{2\pi}\right)\right)\phi'(x), b\dot{\phi} + \left(a +
b\left(\frac{\theta}{2\pi}\right)\right)\phi'(x)\}\nonumber\\
&=& -b^2\left(\frac{\theta}{\pi}\right)\partial_x \delta(x-y) + 2b\left(a +
b\left(\frac{\theta}{2\pi}\right)\right)\partial_x \delta(x-y)
\end{eqnarray}
where the map (\ref{m20}) has been used as well as the brackets satisfied by $\phi$.
We immediately see that we can obtain what we intended for if we make
$a=-b\left(\frac{\theta}{2\pi}\right)$ and $b=1$. This particular map takes the
hamiltonian (\ref{d3}) to
\begin{eqnarray}
\label{m23} H \rightarrow \int dx \left[ \frac{\Sigma'^2}{2} + \frac{P^2}{2} \right]
\end{eqnarray}
where $P = \dot{\Sigma} + \frac{\theta}{\pi}\Sigma'$ is the canonical momentum as
can be explicitly verified calculating the remaining bracket structure:
\begin{eqnarray}
\label{m24} \{\Sigma'(x),P(y)\} &=& \partial_x \delta(x-y)\nonumber\\
\{P(x),P(y)\} &=& 0
\end{eqnarray}
which along with the hamiltonian (\ref{m23}) and the brackets (\ref{m22})
\begin{eqnarray}
\label{m25} \{\Sigma'(x),\Sigma'(y)\} = -\left(\frac{\theta}{\pi}\right)\partial_x
\delta(x-y)
\end{eqnarray}
defines the dual formulation of (\ref{d3}), (\ref{d4}).


Finally we would like to discuss the existence of a duality between the IR and UV
scales as defined in a general way by the quantities $B^k$ and $\Theta^k$ in the
introductory remarks of our results (\ref{c02}). In the model we studied in this
work there are two relevant constants: $g$ with mass dimension 1 and $\theta$ which
is dimensionless. As discussed previously, the role of $g$ is to set the scale of
our approximations related to the Goldstone-Wilczek mechanism thus defining the
region of validity of the effective scalar theory. $\theta$ on the other hand
defines the deformation of the canonical structure but being dimensionless it seems
to lack the significance that was carried by its analogs $B^k$ and $\Theta^k$. But
we should be careful here because the interplay of $g$ and $\theta$ in the
definition of scale and deformation makes this analysis non-trivial.

To make a proper analysis it is interesting to make a Fourier decomposition of the
effective scalar theory and study the duality mode by mode. We shall adopt the
$O(2)$ decomposition, instead of the usual textbook $U(1)$, given by (for details
see \cite{banerjee-wotzasek1, banerjee-wotzasek2}).
\begin{eqnarray}
\label{c03} \phi(x,t) = \int dk\; q_a(t, k) \hat{e}_a(k,x); \;\;\;\;\; a,b=1,2
\end{eqnarray}
where the $O(2)$ basis is such that
\begin{eqnarray}
\label{c04}
\int dx\; \hat{e}_a(k, x) \hat{e}_b(k',x)= \delta_{ab}\delta(k-k').
\end{eqnarray}
Using this and the property: $\partial_x \hat{e}_a(k,x') =
\varepsilon_{ab}k\hat{e}_b(k,x')$, we find that each mode is controlled by the
Lagrangean
\begin{eqnarray}
\label{c05}  L = \frac12 \dot{q}_a^2 - \frac{k^2}{2}q_a^2 + \theta k
q_a\varepsilon_{ab}\dot{q}_b
\end{eqnarray}
representing a two dimensional harmonic oscillator under the influence of an
external magnetic field $B\sim\theta k$.

In this illuminating mechanical language the duality discussed above becomes
expressed in terms of a two-dimensional harmonic oscillator hamiltonian
\begin{eqnarray}
\label{c06}  H(p,q) = \frac12 p_a^2 + \frac{k^2}{2}q_a^2
\end{eqnarray}
with deformed canonical brackets
\begin{eqnarray}
\label{c07}  \{q_a, q_b\} &=& 0\nonumber\\
 \{q_a, p_b\} &=& \delta_{ab}\delta(k-k')\nonumber\\
 \{p_a, p_b\} &=& k\theta\varepsilon_{ab}\delta(k-k')
\end{eqnarray}
and a dual formulation given by
\begin{eqnarray}
\label{c08}  \ast H = H(\ast p, \ast q)
\end{eqnarray}
\begin{eqnarray}
\label{c09}  \{ \ast q_a, \ast q_b\} &=& \frac{\theta}{k}\varepsilon_{ab}\delta(k-k')\nonumber\\
 \{\ast q_a,\ast p_b\} &=& \delta_{ab}\delta(k-k')\nonumber\\
 \{\ast p_a,\ast p_b\} &=& 0.
\end{eqnarray}
We may therefore define the effective deformation parameters
\begin{eqnarray}
\label{c10}  \Theta = k\theta
\end{eqnarray}
and
\begin{eqnarray}
\label{c11}  B = \frac{\theta}{k}
\end{eqnarray}
such that $\Theta$ has mass dimension $1$ and $B$ has mass dimension $-1$ as
expected. We are then led to the conclusion that duality may provide a map between
the infra-red scale characterized by $\Theta$ and the ultra-violet scale
characterized by $B$. This self-duality would then tells us that those scales
sustain the same physics in opposition to expected arguments \cite{carmona}. On the
other hand this may give us a hint on the connections of the results found in
\cite{gamboasarrion} and \cite{gamboapoly} regarding the existence of a Lorentz
violation spectrum.

We should be careful though. We are working with a Fourier slice and in the field
theory we must sum over all momentum contributions. In fact, after we sum up all
modes, there should remain no difference between the infra-red and ultra-violet
regions as this scalar theory is scale invariant. But there is a catch. We are
working with an effective theory with a built in scale defined by the coupling $g$.
Even though it does not appear directly in the action it will appear in the
summation of the modes. The scale invariance is lost and $g$ is the defining scale.
The effective theory is valid for $k\ll g$ and the original theory has a Lorentz
violating deformation given by the effective parameter $k\theta$. In order for this
to be a tiny Lorentz violation $\theta$ must be a small quantity such that
$k\theta\ll k$, that is, the scale associated with the Lorentz violation should be
much lower then the scale of the relevant phenomenon, this is why it is considered
an IR deformation. This just amounts to demand that $\theta \ll 1$. Observe that
this demand is sufficient to guarantee that Lorentz violations effects are small
even in the complete theory were fermionic excitations must be considered, as
depicted in the energy scale diagram below.
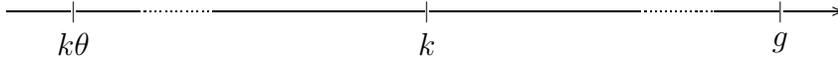
\begin{eqnarray}
 \xymatrix@1{\ar@{-}[rr]_*+++++{k\theta}|| & &\ar@{.}[rr] &\ar@{-}[rr]_*+++++{k}||&  &\ar@{.}[rr] &\ar@{->}[rr]_*+++++{g}||  & &\nonumber\\
          & & & & \txt{The arrow is pointing\\ to increasing energies} & & & &}
\end{eqnarray}

Through duality however the Lorentz violation shows up in the effective parameter
$\frac{\theta}{k}$, this can still be a tiny violation, an UV one due to its mass
dimension, and in this sense there is a map UV/IR. But the condition discussed above
($\theta\ll 1$) does not guarantee anymore that the Lorentz violation is a small
effect. The condition in the dual picture reads $\frac{\theta}{k} \ll \frac 1k$ but
since we also have $\frac 1g \ll \frac 1k$ the only way to be certain that the
Lorentz violating effects are small (even for the complete theory) is to demand that
$\frac{\theta}{k} \ll \frac 1g$ (see the distance scale diagram below) but this does
not follow from the original theory.
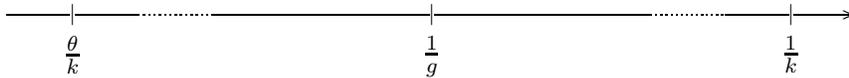
\begin{eqnarray}
 \xymatrix@1{\ar@{-}[rr]_*+++++{\frac{\theta}{k}}|| & &\ar@{.}[rr] &\ar@{-}[rr]_*+++++{\frac{1}{g}}||&  &\ar@{.}[rr] &\ar@{->}[rr]_*+++++{\frac{1}{k}}||  &  &\nonumber\\
        & & & & \txt{The arrow is pointing\\ to increasing distances} & & & &}
\end{eqnarray}

\section{Conclusions and perspectives}

In this work we have put forward another example of a Lorentz violating fermionic
interaction that can be written as an effective theory where all the information
about the Lorentz violation is carried by the deformed canonical relations (or
equivalently, a deformed sympletic sector in the lagrangean). This is an important
result because it helps to corroborate a general expectation that the physical
meaning of these, otherwise \emph{ad hoc}, deformations seems to be an underlying
Lorentz violating dynamics.

We further analyzed this canonical structure making use of duality relations. This
revealed that the deformation appearing in the field momentum sector of the
symplectic matrix can be mapped to the field configuration sector. In both of the
deformations the breaking of Lorentz invariance manifests itself: for the field
momentum sector deformation it shows up as an infra-red effect, for the dual field
configuration deformation it is an ultra-violet effect. But there is also the
possibility that the mapping connects small, yet inaccessible, scales with possibly
observable ones. This may rise an interesting conundrum as it suggests the
possibility that even if we try to hide the Lorentz violation in some yet
unreachable scale it may come to haunt us by duality showing up in the measuring of
some dual observable. We have drawn this conclusion from a particular $(1+1)D$
example but since it was based on general duality properties we think that there are
grounds to believe that it may be a more general feature.

\section{Appendix}

Here we will sketch how the mean value of the fermionic current can be obtained
following the method presented in \cite{af}. We will also comment on how the
effective action (\ref{m07}) can be seen to arise from the bosonized representation
of (\ref{m01}) under the appropriate limits.

The general expression for the mean value of the current is given by the equation
3.12 in \cite{af}:
\begin{eqnarray}
\label{app-indcurr}
 <j^{\mu}(x)> = -i \left[Tr\frac{1}{\pslash-M_0}\gamma^{\mu}\delta(x-y)  +
Tr\frac{1}{\pslash-M_0}\tilde{M}\frac{1}{\pslash-M_0}\gamma^{\mu}\delta(x-y) + ...
\right].
\end{eqnarray}
In the present case
\begin{eqnarray}
\label{app-mass}
 M(\phi) &=& \theta\phi'\gamma^{1} + ge^{\gamma^{5}\phi}\\
 M_0 &=& M(\phi_0) = M(\phi(x_0))\\
 \tilde{M}(\phi) &=& M(\phi) - M(\phi_0),
\end{eqnarray}
where $\phi(x_0)$ is the value of the field $\phi$ evaluated at an arbitrary point
$x_0$. It is convenient to write (\ref{app-mass}) as:
\begin{eqnarray}
\label{app-mass2}
 M(\phi) &=& \theta\phi'\gamma^{1} + g(\cos\phi + \gamma^{5}\sin\phi) =
 \theta\phi'\gamma^{1} + g(\phi_{1} + \gamma^{5}\phi_{2})
\end{eqnarray}
In (\ref{app-indcurr}) the $Tr$ symbol stands for $\gamma$-matrices traces, momentum
integration and space-time integration as well. The omitted higher order terms
refers to higher derivatives. The heart of the matter, as explained in \cite{af}, is
that $\tilde{M}$ contains functions of $x$ so it does not comute with the momenta.
Because of this we must order each term in this expression isolating the $x$'s from
the $p$'s ending with something like
\begin{eqnarray}
\label{app-indcurr2}
 <j^{\mu}(x)> = f^{\mu}(x) \int d^2p g(p)
\end{eqnarray}
This is accomplished using relations like
\begin{eqnarray}
\label{app-comrel}
 \phi\frac{1}{p^2-g^2} &=& \frac{1}{p^2-g^2}\phi + \frac{1}{(p^2-g^2)^2}\left[p^2,\phi\right] +
 higher\;derivatives\nonumber\\
 \left[p^2, \phi(x)\right] &=& \square\phi + 2ip^{\mu}\partial_{\mu}\phi\nonumber\\
 \left[p^{\mu}, \phi\right] &=& i \partial^{\mu}\phi
\end{eqnarray}
This ordering procedure constitutes the core of the calculation, once it is done the
momenta integrations gives only a constant factor.

It is straightforward to see that the first term in (\ref{app-indcurr}) is zero. For
the second term we have
\begin{eqnarray}
\label{app-trace}
 Tr\frac{1}{\pslash-M_0}\tilde{M}\frac{1}{\pslash-M_0}\gamma^{\mu}\delta(x-y) =
 Tr\frac{\pslash+M^{\dagger}_0}{p^2 -g^2}\tilde{M}\frac{\pslash+M^{\dagger}_0}{p^2 -g^2}\gamma^{\mu}\delta(x-y)\nonumber\\
=Tr_y Tr_p \left[\frac{1}{p^2 -g^2}Tr_{\gamma}
\left[(\pslash-M_0)\tilde{M}(\pslash-M_0)\gamma^{\mu}\right]\frac{1}{p^2
-g^2}\right]\delta(x-y)
\end{eqnarray}
We can first perform the $\gamma$-trace. Further neglecting the $\theta$-terms,
which would give a high order $\theta^2$ contribution upon substituting back in the
action, we obtain
\begin{eqnarray}
\label{app-indcurr3}
 <j^{\mu}(x)> = -i Tr_y Tr_p\left[\frac{1}{p^2 -g^2}\left(2g^2p^{\mu}(\tilde{\phi}_1 \phi^0_1 +
 \tilde{\phi}_2 \phi^0_2) + 2ig^2 \varepsilon^{\mu\nu}p_{\nu} (\tilde{\phi}_2 \phi^0_1 -
 \tilde{\phi}_1 \phi^0_2)\right. \right. \nonumber\\
  + \left. \left. 2g^2 (\tilde{\phi}_1 \phi^0_1 +
 \tilde{\phi}_2 \phi^0_2)p^{\mu} - 2ig^2 (\tilde{\phi}_2 \phi^0_1 -
 \tilde{\phi}_1 \phi^0_2)\varepsilon^{\mu\nu}p_{\nu}\right)\frac{1}{p^2
-g^2}\right]\delta(x-y).
\end{eqnarray}
Before performing the momenta integrals we must order this expression by bringing
all momenta to the left, say. After doing that the space-time integral is trivially
evaluated and most momenta integrals results to be null by symmetry. We are left
with
\begin{eqnarray}
\label{app-indcurr4}
 <j^{\mu}(x)> = -i \left[ \int d^2p\frac{1}{(p^2 -g^2)^2}\left(-2ig^2(\phi^0_1\partial^{\mu}\tilde{\phi}_1  +
 \phi^0_2\partial^{\mu}\tilde{\phi}_2 )\right)  \right. \nonumber\\
 + \int d^2p\frac{p^{\mu}p^{\nu}}{(p^2 -g^2)^3}\left(8ig^2 (\phi^0_1\partial_{\nu}\tilde{\phi}_1  +
 \phi^0_2\partial_{\nu}\tilde{\phi}_2 )\right)  \nonumber\\
  + \left. \int d^2p\frac{1}{(p^2 -g^2)^2}\left( - 2g^2 \varepsilon^{\mu\nu}
   \partial_{\nu} \phi \right)\right].
\end{eqnarray}
The first two terms can be seen to cancel each other after doing the momenta
integrals. So the final answer in the approximations we are considering is the same
as the one obtained by Goldstone and Wilczek \cite{gw}.
\begin{eqnarray}
\label{app-indcurr5}
 <j^{\mu}(x)> = -\frac{1}{2\pi}\varepsilon^{\mu\nu}
   \partial_{\nu} \phi(x).
\end{eqnarray}

There is also a nice, heuristic, way to reach the effective theory (\ref{m07}) by
reasoning with the bosonized version of (\ref{m01}) which is given by \cite{gw}:
\begin{eqnarray}
\label{app-bos}
 S = \int d^2x \left( \frac 12 \partial_{\mu}\phi\partial^{\mu}\phi +
 \frac 12 \partial_{\mu}\chi\partial^{\mu}\chi - \frac{\theta}{\sqrt{\pi}} \phi'\dot{\chi}
 - g\mu\cos(2\sqrt{\pi}\chi-\phi)\right)
\end{eqnarray}
where $\chi$ is the bosonized version of the fermionic field and $\mu$ is an
arbitrary energy scale introduced for dimensional reasons. In the approximations we
are considering the fermionic current is viewed as having its origin totally
determined by the original scalar field $\phi$. Furthermore as discussed after
(\ref{m05}) it has a classical character resembling a given external input. This
prompt us, since we are interested only in the scalar $\phi$ field dynamics, to
neglect the bosonized fermionic kinetic term $\frac 12
\partial_{\mu}\chi\partial^{\mu}\chi$ in (\ref{app-bos}). This is in tune with
considering a large energy gap between the characteristic energies of the scalar
$\phi$ field and of the fermions, that is, $\partial\phi \ll g$. In this limit we
have a situation analogous to the London limit forcing the cosine potencial to a
minimum value, fixing the condition $\chi \rightarrow \frac{1}{2\sqrt{\pi}}\phi$,
after which we obtain the effective theory (\ref{m07})
\begin{eqnarray}
\label{app-effthe}
 S = \int d^2x \left( \frac 12 \partial_{\mu}\phi\partial^{\mu}\phi
  - \frac{\theta}{2\pi} \phi'\dot{\phi}\right)
\end{eqnarray}

\section{Note Added}

It came to our attention, after finishing the first version of this paper, an
interesting work by Passos and Petrov, \cite{petrov}. Working also in $(1+1)D$ they
have obtained another example of the kind of phenomenon studied here: a Lorentz
violating fermionic interaction inducing a deformed scalar theory. Their results
differ from ours since they work with a model containing two scalar fields.
Nevertheless it helps to corroborate our general claim that deformed canonical
structures might be described by fermionic effects.

\section{Acknowledgments}

We thank Prof. J. Gamboa for suggestions in an earlier version of this manuscript
and for fruitful conversations. The authors would also like to thank Funda\c{c}\~ao
de Amparo \`a Pesquisa do Estado do Rio de Janeiro (FAPERJ) and Conselho Nacional de
Desenvolvimento Cient\' ifico e Tecnol\'ogico (CNPq) and CAPES (Brazilian agencies)
for financial support.

\end{document}